\documentstyle[preprint,prl,epsfig,eqsecnum,aps,floats,amssymb]{revtex}
\newcommand{\XX}{\mbox{$\, \times \,$}}
\newcommand{\DP}{\mbox{$\Delta\phi$}}
\newcommand{\DE}{\mbox{$\Delta\eta$}}

\newcommand{\ncal}{\mbox{$n_{\rm{cal}}$}}
\newcommand{\ntrk}{\mbox{$n_{\rm{trk}}$}}

\newcommand{\twoD}{(2D)}
\newcommand{\ttt}{\mbox{$2\!\rightarrow\! 2$}}
\newcommand{\sigsi}{\mbox{$\sigma_{\rm singlet}$}}

\begin{document}
\tightenlines


\title{Probing Hard Color-Singlet Exchange in $p\bar{p}$ Collisions \\
   at $\sqrt{s}$ = 630~GeV and 1800~GeV}

\author{                                                                      
B.~Abbott,$^{40}$                                                             
M.~Abolins,$^{37}$                                                            
V.~Abramov,$^{15}$                                                            
B.S.~Acharya,$^{8}$                                                           
I.~Adam,$^{39}$                                                               
D.L.~Adams,$^{48}$                                                            
M.~Adams,$^{24}$                                                              
S.~Ahn,$^{23}$                                                                
H.~Aihara,$^{17}$                                                             
G.A.~Alves,$^{2}$
% ...
\\                                                                            
\vskip 0.70cm                                                                 
\centerline{(D\O\ Collaboration)}                                             
\vskip 0.70cm                                                                 
}                                                                             
\address{                                                                     
\centerline{$^{1}$Universidad de Buenos Aires, Buenos Aires, Argentina}       
\centerline{$^{2}$LAFEX, Centro Brasileiro de Pesquisas F{\'\i}sicas,         
                  Rio de Janeiro, Brazil}                                     
\centerline{$^{3}$Universidade do Estado do Rio de Janeiro,                   
                  Rio de Janeiro, Brazil}                                     
\centerline{$^{4}$Universidad de los Andes, Bogot\'{a}, Colombia}             
\centerline{$^{5}$DAPNIA/Service de Physique des Particules, CEA, Saclay,     
                  France}                                                     
\centerline{$^{6}$Panjab University, Chandigarh, India}                       
\centerline{$^{7}$Delhi University, Delhi, India}                             
\centerline{$^{8}$Tata Institute of Fundamental Research, Mumbai, India}      
\centerline{$^{9}$Kyungsung University, Pusan, Korea}                         
\centerline{$^{10}$Seoul National University, Seoul, Korea}                   
\centerline{$^{11}$CINVESTAV, Mexico City, Mexico}                            
\centerline{$^{12}$Institute of Nuclear Physics, Krak\'ow, Poland}            
\centerline{$^{13}$Institute for Theoretical and Experimental Physics,        
                   Moscow, Russia}                                            
\centerline{$^{14}$Moscow State University, Moscow, Russia}                   
\centerline{$^{15}$Institute for High Energy Physics, Protvino, Russia}       
\centerline{$^{16}$University of Arizona, Tucson, Arizona 85721}              
\centerline{$^{17}$Lawrence Berkeley National Laboratory and University of    
                   California, Berkeley, California 94720}                    
\centerline{$^{18}$University of California, Davis, California 95616}         
\centerline{$^{19}$University of California, Irvine, California 92697}        
\centerline{$^{20}$University of California, Riverside, California 92521}     
\centerline{$^{21}$Florida State University, Tallahassee, Florida 32306}      
\centerline{$^{22}$University of Hawaii, Honolulu, Hawaii 96822}              
\centerline{$^{23}$Fermi National Accelerator Laboratory, Batavia,            
                   Illinois 60510}                                            
\centerline{$^{24}$University of Illinois at Chicago, Chicago,                
                   Illinois 60607}                                            
\centerline{$^{25}$Northern Illinois University, DeKalb, Illinois 60115}      
\centerline{$^{26}$Northwestern University, Evanston, Illinois 60208}         
\centerline{$^{27}$Indiana University, Bloomington, Indiana 47405}            
\centerline{$^{28}$University of Notre Dame, Notre Dame, Indiana 46556}       
\centerline{$^{29}$Purdue University, West Lafayette, Indiana 47907}          
\centerline{$^{30}$Iowa State University, Ames, Iowa 50011}                   
\centerline{$^{31}$University of Kansas, Lawrence, Kansas 66045}              
\centerline{$^{32}$Louisiana Tech University, Ruston, Louisiana 71272}        
\centerline{$^{33}$University of Maryland, College Park, Maryland 20742}      
\centerline{$^{34}$Boston University, Boston, Massachusetts 02215}            
\centerline{$^{35}$Northeastern University, Boston, Massachusetts 02115}      
\centerline{$^{36}$University of Michigan, Ann Arbor, Michigan 48109}         
\centerline{$^{37}$Michigan State University, East Lansing, Michigan 48824}   
\centerline{$^{38}$University of Nebraska, Lincoln, Nebraska 68588}           
\centerline{$^{39}$Columbia University, New York, New York 10027}             
\centerline{$^{40}$New York University, New York, New York 10003}             
\centerline{$^{41}$University of Rochester, Rochester, New York 14627}        
\centerline{$^{42}$State University of New York, Stony Brook,                 
                   New York 11794}                                            
\centerline{$^{43}$Brookhaven National Laboratory, Upton, New York 11973}     
\centerline{$^{44}$University of Oklahoma, Norman, Oklahoma 73019}            
\centerline{$^{45}$Brown University, Providence, Rhode Island 02912}          
\centerline{$^{46}$University of Texas, Arlington, Texas 76019}               
\centerline{$^{47}$Texas A\&M University, College Station, Texas 77843}       
\centerline{$^{48}$Rice University, Houston, Texas 77005}                     
}                                                                             

\maketitle

\begin{abstract}
We present results on dijet production via hard color-singlet exchange
  in proton-antiproton collisions at $\sqrt{s} = 630$~GeV 
  and 1800~GeV using the D\O\ detector.
The fraction of dijet events produced via color-singlet exchange is measured as
  a function of jet transverse energy, separation in 
  pseudorapidity between the two highest transverse energy jets,
  and proton-antiproton center-of-mass energy.
The results are consistent with a color-singlet fraction
  that increases with an increasing fraction of quark-initiated processes 
  and inconsistent with two-gluon models for the hard color-singlet.
\end{abstract}

\pagebreak

The exchange of a quark or gluon between interacting partons
in hadronic collisions typically 
results in final-state particle production over several
units of rapidity.  In contrast, the exchange of
a color-singlet is expected to yield
a rapidity gap, defined as the absence of particles in a region of rapidity
(or pseudorapidity,
$\eta \equiv -\ln \tan (\theta /2)$, where $\theta$
is the polar angle).
Rapidity gaps are most commonly observed in low
momentum-transfer diffractive and elastic scattering processes, which
are attributed to the exchange of a color-singlet called the pomeron.
%
The observation of 
rapidity gaps between jets  
at both the Fermilab $p\bar{p}$ collider
(Tevatron)~\cite{D01,CDF,D02,CDF2} 
and the DESY $ep$ Collider (HERA)~\cite{Zeus} implies the exchange
of a hard color-singlet.
The measured fraction of dijet events arising from color-singlet 
exchange is roughly $1\%$ in proton-antiproton
collisions~\cite{D01,CDF,D02,CDF2} and $10\%$ in 
positron-proton collisions~\cite{Zeus}.
These rates are   too large to be explained by electroweak boson exchange and
indicate a strong-interaction process~\cite{D02,Zeus}.

The final-state particle multiplicity distribution between
jets is used to distinguish between
color exchange and color-singlet exchange.
By identifying an excess of low multiplicity events using 
the D\O\ calorimeter, tracking system, or both,
we measure the observable fraction of color-singlet exchange in dijet events
(``color-singlet fraction'').
This Letter presents new measurements by the D\O\ 
Collaboration of the 
color-singlet fraction as a function of 
  jet transverse energy ($E_T$),
  pseudorapidity separation between the 
  two highest $E_T$ jets ($\Delta\eta=|\eta_1-\eta_2|$), and proton-antiproton
  center-of-mass energy ($\sqrt{s}$). 
We then compare these 
   measurements to hard color-singlet models.

The D\O\ detector and trigger system are described in Ref.~\cite{d0NIM}.
The data samples were accumulated during the 
  1994--1996 Tevatron run primarily at a  
  proton-antiproton center-of-mass
energy of 1800~GeV, with a  short run at 630~GeV 
  during this period.
Jets were found in the liquid-argon calorimeter (full coverage to
$|\eta|\!<\!4.1$)
using a cone algorithm 
with radius  ${\cal R}\!= 0.7$ in $\eta-\phi$ space,
where 
$\phi$ is the azimuthal angle.
Samples of ``opposite-side'' (OS) jet events 
  ($\eta_1\cdot\eta_2<0$) and ``same-side'' (SS) jet
  events ($\eta_1\cdot\eta_2>0$)
  were collected using triggers similar to those  described
  in Ref.~\cite{D02}. 
The triggers required both jets to have $E_T$ above a
threshold which varied from 12 to 25 GeV depending on the trigger.
All triggers required at least two jets with $|\eta|\!>\!1.6$, and 
  the opposite-side triggers required a dijet pseudorapidity 
  separation of $\Delta\eta\!>\!4.0$ at 
$\sqrt{s} = 1800$~GeV and $\Delta\eta\! >\! 3.2$ 
  at 630~GeV.

In the offline analysis, 
the vertex was required to have a longitudinal position 
within 50~cm of the center of the detector
and the two highest transverse energy (leading)
jets were required to have $|\eta|\!>\!1.9$.
The opposite-side samples were required to have 
$\Delta\eta\!>\! 4.0$ between the leading two jets.
Since multiple interactions (more than one proton-antiproton interaction
in the same bunch crossing) may obscure the rapidity gap signal of color-singlet
exchange, we retained only events with a single reconstructed vertex and 
additionally required 
the beam-beam hodoscope timing information 
to be consistent with a single interaction.

Table~\ref{tab:triggers} summarizes the data samples used in this analysis,
including the cut on the transverse energy of the second leading jet
($E_{T2}$), the integrated luminosity, 
and the approximate number of events for each trigger.
At $\sqrt{s} = 1800$~GeV, three exclusive
   opposite-side data samples corresponding to triggers
with low, medium, and high $E_T$ thresholds (``low-$E_T$'', ``med-$E_T$'',
  and ``high-$E_T$'') are used 
  to determine the $E_T$ dependence of the color-singlet fraction.
The $\Delta\eta$ dependence of the color-singlet fraction is  
  measured using the low-$E_T$ and high-$E_T$ samples.
The large statistics high-$E_T$ sample gives the most accurate
  measure of the color-singlet fraction.

\begin{figure}[ht]
\centerline{\epsfig{figure=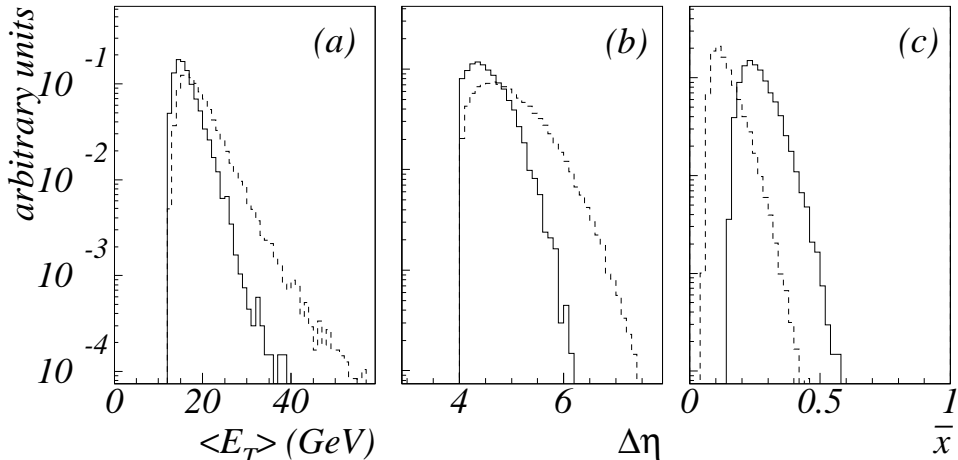,width=5in}}
\caption{Jet characteristics of the 630-OS (solid line) and 1800-OS 
(dashed line) 
  data samples.
The normalized distributions are shown 
for (a) the average $E_T$, (b) 
  $\Delta\eta$, and (c) 
  $\bar{x}$ of the two leading
  jets. 
\label{f:char}
}
\end{figure}

Samples with the same jet $E_T$ threshold of 12~GeV are used to compare
  the $\sqrt{s}=630$~GeV and 1800~GeV color-singlet fractions. 
  This lower $E_{T2}$ cut 
  provides adequate statistics for the 
  630~GeV sample.
Figures~\ref{f:char}(a) and (b) show the average jet $E_T$
and $\Delta\eta$, respectively, 
 of the two leading jets for the 630-OS and 1800-OS samples.
As expected from kinematic considerations, the jets in the 
$\sqrt{s}=1800$~GeV sample reach a higher
$E_T$ and $\Delta\eta$ than the 630~GeV sample.
Figure~\ref{f:char}(c) shows the 
average parton $x$ ($\bar{x}$) for the two samples, where we calculate
$\bar{x}$ using a leading order approximation
($\bar{x}= (x_{+}\! +\! x_{-})/2$;
$x_{\pm} = \sum_{i=1}^{2} E_{Ti}{\rm e}^{\pm\eta_i}/\sqrt{s}$).

Table~\ref{tab:triggers} also lists the  same-side samples at
  630~GeV and 1800~GeV.
Same-side events are useful since they have similar characteristics 
  to opposite-side events but have 
  no observable color-singlet component~\cite{D02}.


\begin{table}[htb]
\vspace{-.1in}
\begin{center}
\begin{tabular}{|l||l|l|l|l|}
Name & $\sqrt{s}$ (GeV)  & $E_{T2}$ (GeV) & Luminosity (nb$^{-1}$)  &
Events \\ \hline \hline
 low-$E_T$  &1800 & $15-25$ & 86 & 27000  \\
 med-$E_T$  &     & $25-30$ & 1300 & 21000  \\
 high-$E_T$ &     & $>30$ & 11000 & 72000 \\
 1800-OS    &     & $>12$ & 86 & 48000  \\
 1800-SS    &     &  $>12$ & 18 & 57000  \\ \hline
 630-OS     & 630 & $>12$ & 520 & 6700  \\
 630-SS     &     & $>12$ & 350 & 6800
\end{tabular}
\end{center}
\vspace{-.15in}
\caption{Data samples showing the offline cut on
the transverse energy of the second leading jet
($E_{T2}$), integrated luminosity,
and approximate number of events after all offline cuts.
All triggers are opposite-side jet triggers
except for 1800-SS and 630-SS, which are same-side
jet triggers.
\label{tab:triggers}}
\end{table}

We determine the amount of color-singlet exchange 
in the opposite-side samples by utilizing the multiplicity distributions
measured by the calorimeter and the tracking system in the
pseudorapidity region $|\eta|\!<\!1$. 
This region has little sensitivity to contamination from the leading jets,
which have a minimum $|\eta|$ of 1.9.
We measure the multiplicity in the electromagnetic calorimeter by counting
the number of towers ($\DE \XX \DP = 0.1 \XX 0.1$)
with $E_T\!>\!200$~MeV (\ncal).  
The electromagnetic section of the calorimeter has
a low level of noise and the ability to detect both charged and neutral
particles. We also count the number 
of tracks in the central drift chamber (\ntrk), which is
a non-magnetic tracking system that detects charged particles to
low momenta and provides an independent measure of the multiplicity.

\begin{figure}[hbt]
\centerline{\epsfig{figure=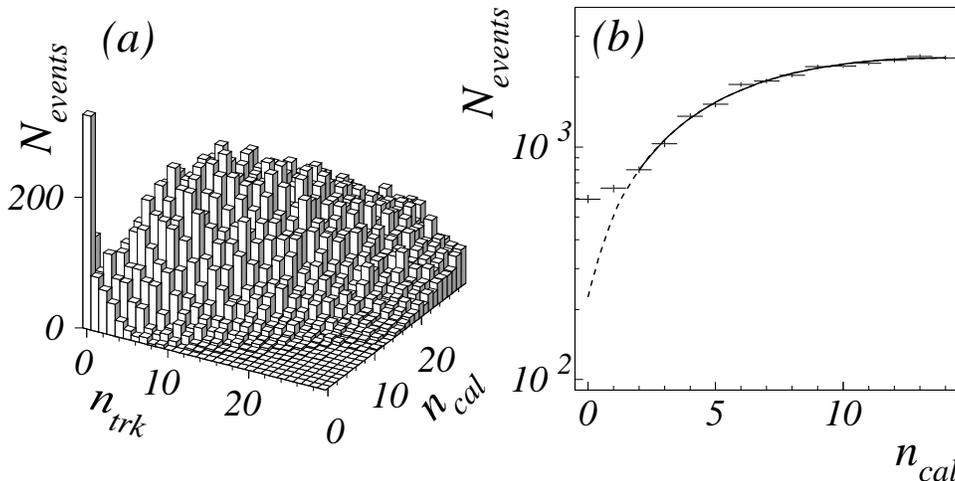,width=5in}}
\caption{ Multiplicity in the region $|\eta|\!<\!1$
between the two leading jets for the high-$E_T$ 
sample: (a) 
  two-dimensional multiplicity, \ncal\ vs. \ntrk; (b)
\ncal\  only with NBD fit. 
\label{f:1800}
}
\end{figure}

Figure~\ref{f:1800}(a) displays the \ncal\ versus \ntrk\ distribution in the 
central region ($|\eta|\!<\!1$)  between the jets for the high-$E_T$ 
($\sqrt{s} = 1800$~GeV)
sample. A large excess of events is observed at low multiplicity, 
consistent with the expectations of a color-singlet exchange signal
and  similar to that observed previously~\cite{D02,CDF2}.
Figure~\ref{f:1800}(b) shows the \ncal\ distribution
  with a three-parameter negative binomial distribution (NBD) fit, 
which is used 
  to determine the color-exchange background.
NBD's have previously been shown to provide a good description
of particle multiplicity distributions 
  in proton-antiproton collisions~\cite{D02}.

We measure the color-singlet fraction ($f_S$) by dividing
the number of events above
the  NBD fit for $\ncal \!\leq\!1$ 
by the total number of events in the sample.
This quantity is insensitive to noise: after the suppression
of a few known noisy calorimeter towers,
less than $1\%$ of 
events taken during no-beam runs
have more than one tower above threshold.
In addition, 
the color-singlet fraction
is not sensitive to jet finding and trigger efficiencies, which 
cancel in the ratio
since the jets found in color-singlet events are indistinguishable
from the jets in color-exchange events.

To minimize sensitivity to residual contamination from 
  multiple proton-antiproton interactions, 
  we fit  to the low multiplicity 
  region of the data.
The fits start where
the signal to background ratio
 becomes negligible ($\ncal\!=\! 2$)
  and  end near the maximum of the multiplicity distribution 
 ($\ncal\!=\!11$ at 630~GeV and $\ncal\!=\!14$ at 1800~GeV).
The fits are then extrapolated to zero multiplicity to
determine the background.
This method was tested by generating and fitting ensembles of 
  multiplicity distributions with and without a known signal 
  component at low multiplicity.
These studies confirm that the fitting method is unbiased 
  and also provide a measure of the systematic error~\cite{Tracy,Jill}.
An additional uncertainty due to the choice of the
 starting multiplicity bin used for the fit is estimated by
 determining the background using a starting bin of
 $\ncal\!=\!3$; the resulting change in background extrapolation 
 is applied as an additional systematic error.

Corrections are applied to $f_S$ to account for 
residual multiple interaction contamination in the samples as well as
for  single-vertex events erroneously tagged as multiple interaction
events. 
The extra multiplicity 
from secondary interactions causes color-singlet events
to be erroneously measured 
with a large multiplicity.
By studying the shape of the multiplicity distribution as a function of 
luminosity, a correction is determined that results in roughly
a 30\% increase in $f_S$. The vertex correction 
reduces $f_S$ by about 10\% since 
color-singlet events always have one vertex, but about 10\% of color
exchange events with one interaction have two reconstructed 
vertices~\cite{vtx}.

Applying the fitting method to the $\sqrt{s} = 1800$~GeV 
high-$E_T$ data ($E_{T2}\!>\!30$~GeV)
   shown in Fig.~\ref{f:1800}(b), we obtain
  $f_S = [0.94\pm0.04$(stat)$\pm 0.12$(syst)]\%.
The systematic error
is dominated by the uncertainty in the background
  subtraction ($0.09\%$) but also includes uncertainties in the 
  single interaction determination ($0.07\%$), offline cuts ($0.03\%$), and 
  jet energy scale and reconstruction ($0.02\%$).
This measurement is the most precise to date and is consistent with 
  previously published measurements
  at the Fermilab Tevatron with similar minimum jet 
$E_T$~\cite{D01,CDF,D02,CDF2}.

The value of $f_S$ does not depend on the exact method used to measure  
particle multiplicity. We vary the 
calorimeter tower threshold
used for determining \ncal\  
between 150~MeV and  350~MeV 
 (the nominal value is 200~MeV) and obtain a consistent value for the fraction.
Redefining multiplicity using a ``cluster'' of
neighboring calorimeter towers 
to account for possible calorimeter showering effects, or
using the tracking  multiplicity (\ntrk) instead of \ncal\
results in a measured color-singlet fraction within a few
per cent of the nominal value.

\begin{figure}[htb]
\centerline{\epsfig{figure=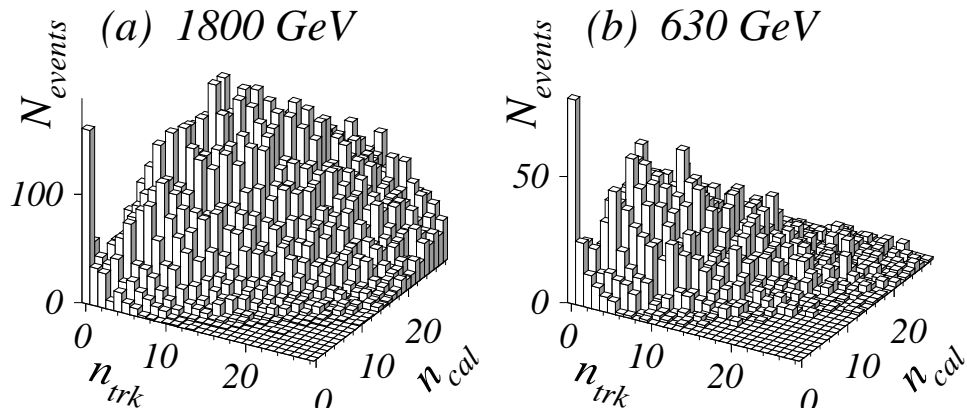,width=5in}}
\caption{ Two-dimensional multiplicity (\ncal\ vs. \ntrk) 
in the region $|\eta|\!<\!1$
for the (a) 1800-OS and (b) 630-OS
samples.
\label{f:mult}
}
\end{figure}

Figure~\ref{f:mult}(a) shows the multiplicity distribution  
   between the jets for the 1800-OS sample with $E_{T2}\!>\!12$~GeV.
Fitting the \ncal\ distribution and measuring the color-singlet 
  fraction, we obtain $f_S = [0.54\pm0.06$(stat)$\pm 0.16$(syst)]$\%$.
This value is about $2\sigma$
smaller than the value measured for $E_{T2}\!>\!30$~GeV,
  indicating a dependence of the color-singlet fraction on jet $E_T$.

Figure~\ref{f:mult}(b) shows the multiplicity distribution
  between the jets for the 630-OS sample.
For this sample we obtain 
  $f_S = [1.85\pm0.09$(stat)$\pm 0.37$(syst)]$\%$.
The uncertainties in the 630-OS and 1800-OS measurements are dominated
by the fit errors, which are not correlated between the two samples.
We calculate the significance of the difference between 
the two values as $3.1\sigma$ in the limiting case
that the uncertainties are completely uncorrelated.

The ratio of the two measurements for $E_{T2}\!>\!12$~GeV is 
  $R_{1800}^{630}\! \equiv \! f_S(630)/f_S(1800) \! = \! 3.4\pm1.2$.
We verify that the lower average multiplicity for 630~GeV does not
  affect the measured ratio by raising
the electromagnetic tower threshold of the 1800~GeV data
  so that the leading 
  edge of the \ncal\ distribution matches that of the 630~GeV sample.
We also obtain
a consistent value for the ratio when we measure it 
using a two-dimensional
\ncal\ versus \ntrk\ multiplicity method (described below).

Measuring the color-singlet fraction as a function of $E_T$
and $\Delta\eta$ requires splitting the data into
several bins. The NBD fitting uncertainty becomes unacceptably large
for these smaller statistics sub-samples, so
a new two-dimensional multiplicity method is employed
to reduce the uncertainty.
For each bin of $E_T$ or $\Delta\eta$,
we first determine the fraction of events 
with $\ncal + \ntrk\! \leq\! 1$.
We then obtain the ``2D'' color-singlet fraction
($f^{\twoD}_{S}$) by
subtracting the appropriate color-exchange background 
and correcting by the acceptance relative to the NBD
method.

The color-exchange background for the ``2D'' fraction
  is determined using the following procedure.
The NBD fit to the calorimeter multiplicity distribution for
  each full sample is used to determine
  the total background  ($\ncal \!\leq\!1$) for that sample.
This ``1D'' background is converted to a
  ``2D'' background by multiplying
  by the ratio $N(\ncal + \ntrk \!\leq\!1)/N(\ncal \!\leq\!1)$ 
  obtained from the
  same-side multiplicity distribution in the same detector region.
The same-side data are thus used only to account for correlations 
between \ncal\ and \ntrk.
The average background is roughly  25\% for each   sample.
The dependence of the background on $E_T$ 
and $\Delta\eta$  is 
measured using a low-multiplicity  control region 
($3\!\leq\!\ncal + \ntrk\!\leq\!5$)
where the color-singlet contribution is negligible.
The final background for each sample is determined by multiplying the
  average background  with its $E_T$ and $\Delta\eta$ dependence. 

The background is then subtracted from each 
$E_T$ and $\Delta\eta$ bin of the uncorrected ``2D'' 
fraction. The background has little dependence on 
$E_T$ and $\Delta\eta$, so the dependence of the color-singlet fraction 
on these variables before 
and after background subtraction is quite similar. 

Comparing the color-singlet fraction from the ``2D'' method to 
$f_S$ from the NBD fitting method for each sample, we determine 
that the ``2D'' method has a relative
acceptance of $80$\% for color-singlet events. This acceptance
has been determined to have little dependence on
$E_T$ and $\Delta\eta$, and its uncertainty
(3\% over the $E_T$ range and 6\% over the $\Delta\eta$
range) is included in subsequent shape comparisons to 
different color-singlet models.
We obtain the final color-singlet fractions 
$f^{\twoD}_{S}(E_T)$ and $f^{\twoD}_{S}(\Delta\eta)$ by
correcting for the color-singlet acceptance determined from the 
high-$E_T$ sample. 

\begin{figure}[htb]
\centerline{\epsfig{figure=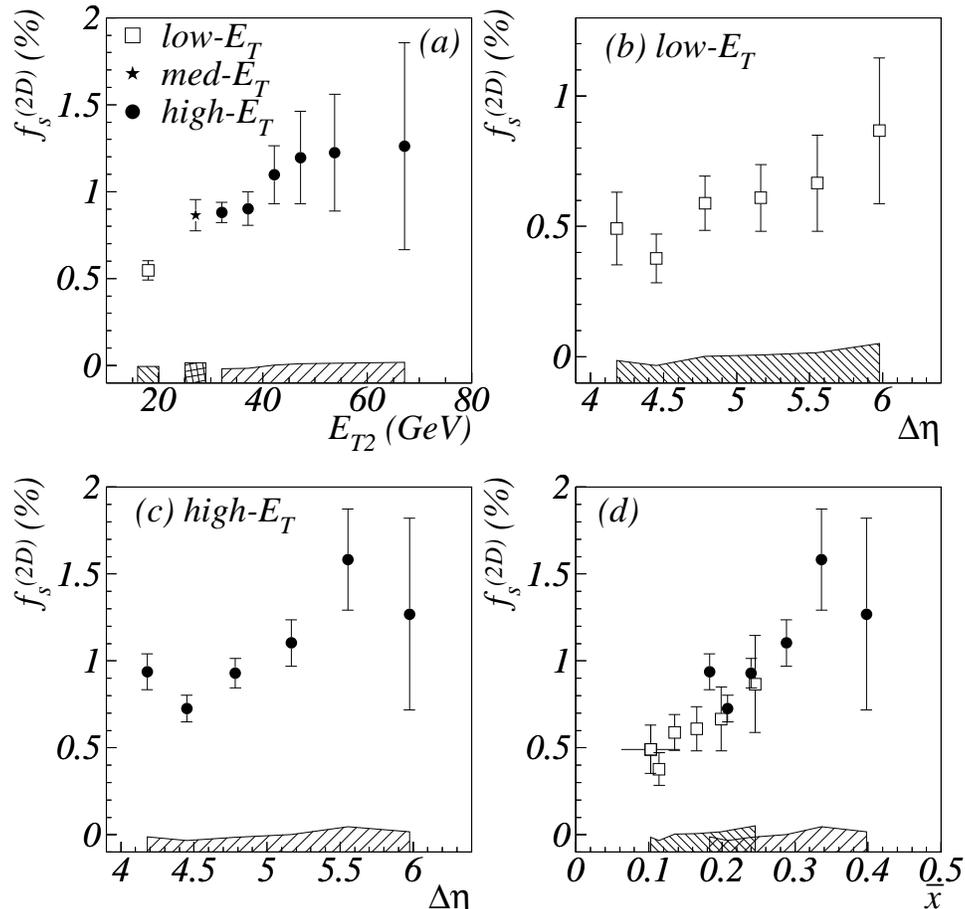,width=5in}}
\caption{The color-singlet fraction $f_S^{\twoD}$: (a)
as a function of the second leading jet $E_T$; 
 as a function of $\Delta\eta$  
between the two leading jets for (b) the low-$E_T$ sample and (c) 
the high-$E_T$ sample;
(d) as a function of  $\bar{x}$ 
for each 
 $\Delta\eta$ bin in (b) and (c). The horizontal error bar on the first 
$\bar{x}$ bin
 shows the $1\sigma$ spread in $\bar{x}$ values entering that (representative)
 bin.
Statistical error bars and relative 
 normalization uncertainties for each sample 
 (hatched bands) are shown.
\label{f:et}
}
\end{figure}

Figure~\ref{f:et}(a) shows the color-singlet fraction $f_S^{\twoD}(E_T)$
  at 1800~GeV obtained using the
  low-$E_T$, med-$E_T$, and high-$E_T$ data samples. 
The error bars show the statistical uncertainty,
  while the bands show the relative normalization
  uncertainties between the different data samples.
We bin by second leading jet $E_T$ ($E_{T2}$) 
  to mimic the way the dijet triggers select events.
In contrast, binning by average dijet $E_T$ 
  selects primarily back-to-back dijet events near the trigger $E_T$ threshold.
Since color-singlet events are more likely to be back-to-back
  than color-exchange events, using an
  average  $E_T$ method would bias the color-singlet 
fraction as a function of $E_T$.

Figures~\ref{f:et}(b) and \ref{f:et}(c) show the color-singlet fraction 
  $f_S^{\twoD}(\Delta\eta)$
  at 1800 GeV for the low-$E_T$ and high-$E_T$ samples, respectively.
The error bars show the statistical uncertainty,
  while the bands show the relative normalization
  uncertainties between the different data samples.

Figures~\ref{f:et}(a)--(c) show 
  that the measured 
  color-singlet fraction tends to increase with $E_T$ and $\Delta\eta$, 
  implying that the fraction increases with parton $x$.
Replotting Fig.~\ref{f:et}(b)--(c) in Fig.~\ref{f:et}(d) as a function of 
$\bar{x}$, with $\bar{x}$ averaged over each
$\Delta\eta$ bin,
  shows a rise in the measured color-singlet fraction with $\bar{x}$.
In principle it is possible to measure the color-singlet fraction
directly as a function of parton $x$.
However, 
bin-dependent
acceptance effects related to the dijet trigger make it preferable
to do the analysis in the observables $E_T$ and $\Delta\eta$.

Measuring the color-singlet fraction as a function of $E_T$, $\Delta\eta$, and
  $\sqrt{s}$ probes the nature of hard color-singlet exchange. 
If the color-singlet exchange process were identical to 
  single-gluon exchange 
  except for different coupling factors to quarks and gluons,
  the color-singlet fraction would depend only on parton 
  distribution functions through the parton $x$.
Thus, for a color-singlet that couples more strongly to gluons (quarks),
  the color-singlet fraction would fall (rise) as a function of increasing 
  $x$,  since the gluon distribution becomes relatively suppressed  
  as $x$ increases.
This scenario would imply a decreasing (increasing) color-singlet fraction with
  increasing jet $E_T$ and $\Delta\eta$ or decreasing $\sqrt{s}$.
The observed color-singlet fraction may deviate from this
  simplistic behavior, however, if the dynamics of color-singlet exchange are 
  distinct from the dynamics of single-gluon exchange.
Currently QCD does not 
  account for the existence of hard color-singlet exchange,
  but higher-order QCD processes 
  have been proposed to explain
  this phenomenon~\cite{Dok,BJ,UW,Mueller,Duca,Buch,Halzen}.

The measured color-singlet fraction 
includes the probability that the color-singlet events are 
observable, and can be written
as $f_{S}=S \cdot  \sigsi /\sigma$, where
$\sigsi/\sigma$ is the fraction of dijet events produced by color-singlet
exchange, and $S$ is the survival probability (probability that
the rapidity gap   is not contaminated by particles from spectator
interactions).  
This survival probability ($S\!\sim\! 10\%$) cannot
be measured independently from the color singlet, and
is assumed not to 
depend on the $x$ or type
of the initiating partons in the
hard scattering~\cite{BJ,GLM}.
Although initial estimates of $S$ had a weak ($\sim\! 1/\ln s$) 
  dependence on center-of-mass energy~\cite{GLM},
  a recently revised estimate gives a stronger dependence~\cite{Gotsman}.

We compare our color-singlet results to Monte Carlo color-singlet models using
 {\footnotesize HERWIG}~\cite{HW}, which includes parton showering and 
 hadronization of the final state partons.
Monte Carlo samples were generated using CTEQ2M~\cite{CTEQ}
  parton distribution functions~\cite{pdfsNote}.
As in the data, jets were reconstructed using a cone algorithm
  with a radius of ${\cal R}\!=\!0.7$.
Detector effects were simulated by smearing the jet $E_T$ and $\eta$
  according to measured jet resolutions.
The two leading jets were required to have
  $|\eta|>1.9$ and $\Delta\eta>4.0$ as in the data samples.

For all the models the color-singlet fraction as a function of $E_T$ and 
$\Delta\eta$  is obtained by dividing
the binned color-singlet cross section 
by the corresponding {\footnotesize HERWIG} QCD cross section. 
A rapidity gap requirement was not imposed on any of the Monte Carlo samples
since identifying rapidity gaps
depends on the details of hadronization
and the modeling of spectator interactions.
The rapidity gap survival probability is
assumed to be independent of $E_T$ and $\Delta\eta$, and 
its magnitude is absorbed in an overall normalization factor.

We use a Bayesian probability method to fit the color-singlet fraction
predicted by the Monte Carlo models
independently to the $f^{\twoD}_S(E_T)$ data (Fig.~\ref{f:et}(a))
and the $f^{\twoD}_S(\Delta\eta)$ data (Fig.~\ref{f:et}(b)--(c)).
Although the measurements of the color-singlet fraction as a function
of $E_T$ and $\Delta\eta$ are statistically correlated, they contain
complementary information.
The normalization, which is not well-determined for any of the models,
is the only free parameter in the fits to each model, except for 
the ``free-factor'' model (discussed below).

The corrections, background, acceptance, Monte Carlo expectations, 
and uncertainties in these quantities
are included in the Bayesian integral as prior probability distributions.
Since there has been no previous comparison to color-singlet models,
we assign an identical prior probability for each model and
assume a flat prior probability (no preferred value) for each fit
parameter. We obtain the posterior probabilities for the fit parameters 
by integrating  the likelihood function over 
all prior probability distributions.
For each model, we determine the fit parameter values that
maximize the posterior probability
for that model.
The uncertainty in each fit parameter is obtained from its posterior
probability distribution by determining the 
68\% confidence interval which has equally probable endpoints.
If one of the boundaries is restricted by physical
constraints then we use
the smallest 68\% confidence interval to determine the uncertainty.
The maximum of the likelihood is taken as a measure of the 
relative goodness-of-fit of that model.

The Bayesian method was chosen for fitting the data since
the conventional $\chi^2$ method introduces a normalization bias
in the case of correlated uncertainties~\cite{Dag}.
This bias is small when the model describes the data well, but for 
some of the models the fitted normalization from the 
$\chi^2$ method was as much as a factor of two smaller than the
corresponding Bayesian value.

\begin{figure}[htb]
\centerline{\epsfig{figure=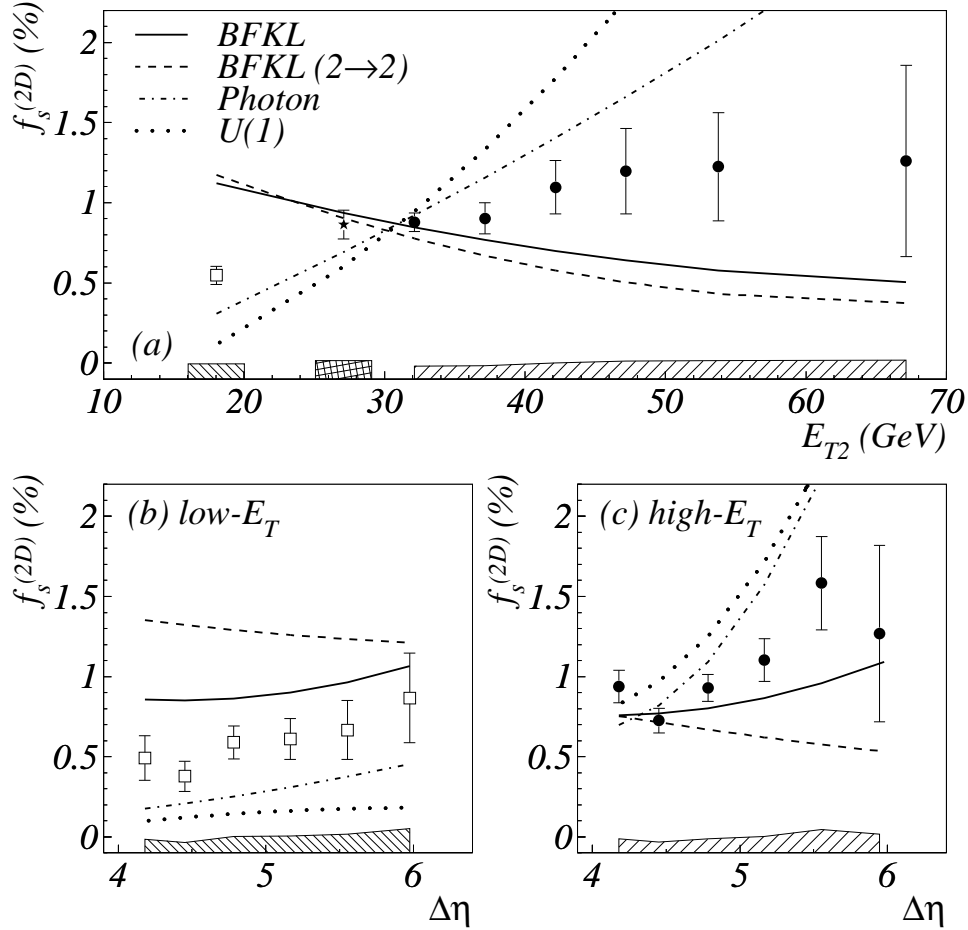,width=5in}}
\caption{Fits of Monte Carlo models 
to the color-singlet fraction (a) $f_S^{\twoD}(E_T)$ 
   and (b)--(c) $f_S^{\twoD}(\Delta\eta)$ for the low-$E_T$ sample and  
  the high-$E_T$ sample, respectively.
Shown are comparisons to BFKL jet level (solid line), BFKL 
$2\!\rightarrow\!2$ 
parton level 
 (dashed line), 
  photon (dot-dashed line), and U(1) (dotted line) models.
\label{f:fithw}
}
\end{figure}

\begin{figure}[htb]
\centerline{\epsfig{figure=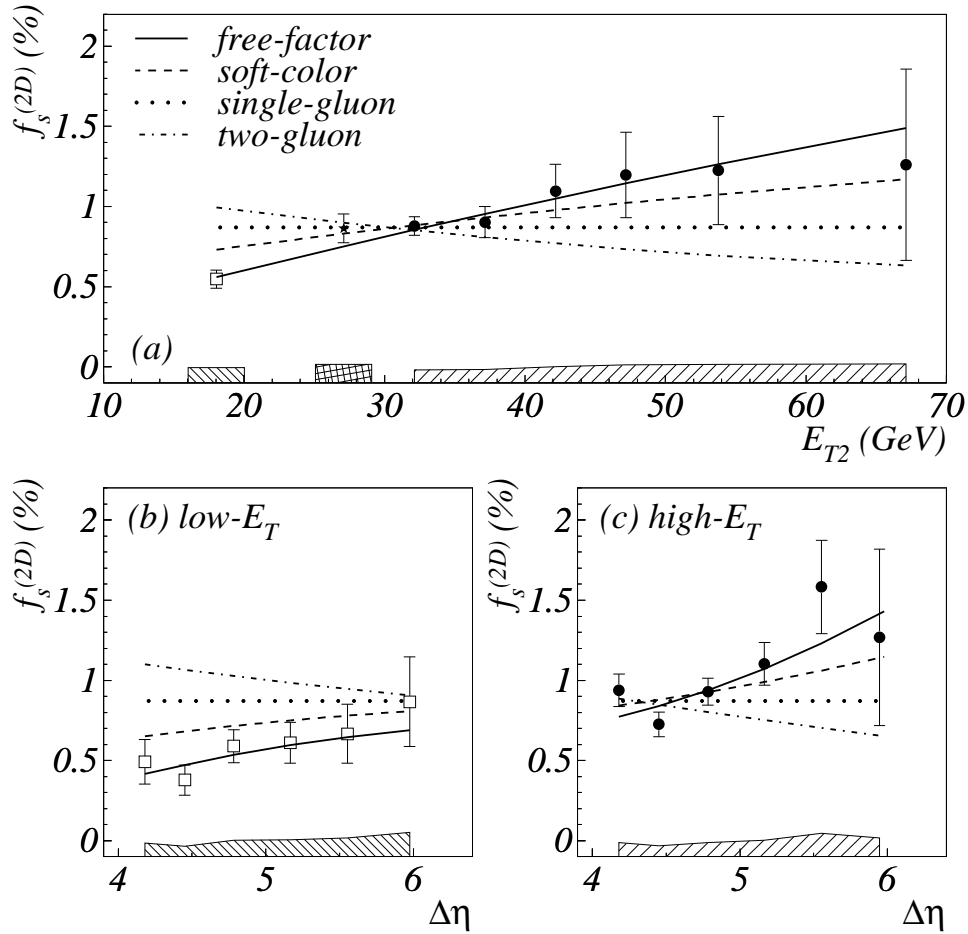,width=5in}}
\caption{Fits of color-factor models to the color-singlet fraction 
(a) $f_S^{\twoD}(E_T)$ 
and (b)--(c) $f_S^{\twoD}(\Delta\eta)$ for the low-$E_T$ sample and  
the high-$E_T$ sample, respectively.
Shown are comparison to free-factor (solid line), 
 soft-color (dashed line), single-gluon (dotted line),
 and simple two-gluon (dot-dashed line)  models.
\label{f:fitnaive}
}
\end{figure}

Figures~\ref{f:fithw}--\ref{f:fitnaive} 
show fits to the measured color-singlet fraction.
Even though the fraction versus $\Delta\eta$ and $E_T$ are fit separately,
the fit parameters obtained from each variable do not differ appreciably
for any given model.
Since the $\Delta\eta$ distributions for 
the low-$E_T$ and high-$E_T$ samples
are fit simultaneously, a model prediction that is above the data for
one plot would be below the data for the other.

The dashed line in Fig.~\ref{f:fithw} (BFKL \ttt) 
shows a parton-level color-singlet prediction
of a two-gluon exchange process incorporating the BFKL
dynamics of Mueller and Tang~\cite{Mueller,MuellerNote}.
The solid line is the BFKL color-singlet fraction 
at the jet level after hadronization.
The BFKL models do not give an adequate description of the data,
primarily due to their prediction of  decreasing  
color-singlet fractions with  $E_T$.
Assuming a survival probability of about 10\%, the normalizations
obtained for the BFKL models are a factor of 6--9 lower than
the data.

Note that the BFKL \ttt\ color-singlet fraction falls with $\Delta\eta$
while the jet-level fraction rises.
Although the BFKL cross section prediction has the same behavior with 
$\Delta\eta$ at both levels,
parton showering for standard QCD processes 
reduces the jet $\Delta\eta$ separation at the jet level relative 
to the parton level.
The resulting BFKL color-singlet fraction prediction is thus sensitive to
higher-order effects in QCD processes,
casting doubt on the reliability of current parton-level predictions.

Although the rate of $t$-channel photon exchange is too small to describe
   D\O\ data~\cite{D02},
   the photon-exchange process in 
 {\footnotesize HERWIG} can be used to
  investigate two models in which the color-singlet couples only to quarks:
  a massless photon-like singlet and a massive U(1) gauge boson~\cite{Carone}.
To simulate the exchange of a massive U(1) gauge boson 
that couples to baryon number, the photon cross section 
  is multiplied by a kinematical factor of
  $(1/(1+m_B^2/E_T^2))^2$ which accounts for the exchange of a boson
  with mass $m_B$.  The coupling $\alpha_B$ of the U(1)
gauge boson is absorbed in the overall normalization.
The behavior of the resulting color-singlet fraction is similar to photon 
exchange except for a steeper rise
as a function of $E_T$.
The lowest allowed mass of $m_B = 20$~GeV$/c^2$~\cite{Carone} 
  gives the best fit to the data.
Figure~\ref{f:fithw} shows the fits to the photon (dot-dashed line) 
  and U(1) (dotted line) models. 
These models do not give an adequate description of the data
  due to their prediction of a steep rise in the color-singlet fraction with 
  $E_T$ and with $\Delta\eta$ for the high $E_T$ sample~\cite{photonNote}. 

In addition to the color-singlet processes in 
 {\footnotesize HERWIG},
  we construct a set of simple  ``color-factor'' models
which explore different possibilities for the coupling of the color
singlet to quarks and gluons. In this approach,
the color-singlet fraction is parametrized 
  as a weighted sum of the relative
  fractions of quarks and gluons from parton distribution functions.
The weights represent effective ``color factors'' that enhance or suppress
  coupling to initial quarks and gluons compared to pure single-gluon exchange.
This parametrization can be written as
\begin{eqnarray*}
  f_S & =  & f_{N}\cdot \lgroup
  C_{qq}\cdot q_p q_{\bar{p}} + C_{qg}\cdot (q_pg_{\bar{p}} + 
  g_pq_{\bar{p}}) + C_{gg}\cdot   g_pg_{\bar{p}} \rgroup
\end{eqnarray*}
where $f_{N}$ is a normalization factor, and $q_i$ and $g_i$ denote the 
  relative fractions of initial quarks and gluons from hadron 
  $i$ $(i = p,\bar{p})$. $C_{qq}$,
  $C_{qg}$, and $C_{gg}$ are the effective color factors for 
  quark-quark, quark-gluon, and gluon-gluon processes,
respectively~\cite{HalzenNote}, where the subscripts refer to
the identity of the interacting partons from the proton and anti-proton.
We apply this formalism to several color-singlet models and
  make quantitative comparisons to the data.
 
The soft-color rearrangement model, which has been used to explain 
  rapidity gaps in positron-proton collisions~\cite{Buch}, 
  has recently been  generalized as an alternative QCD-motivated
  explanation for rapidity gaps in hadron-hadron collisions~\cite{Halzen}.  
In this model, color flow (via the exchange of a single gluon) 
  can be cancelled by the exchange of 
  soft gluons, leading to a rapidity gap and thus
  an effective colorless exchange. 
This cancellation is more likely for 
  initial-state quark processes than for gluon processes since 
  quark states have fewer
  possible color flow configurations.
Based on counting arguments with three quarks and eight gluons, a
reasonable choice of color factors    
  is $C_{qq}\!=\!1/9$, $C_{qg}\!=\!1/24$ and $C_{gg}\!=\!1/64$~\cite{Halzen}.
The ``soft-color'' model predicts a
  color-singlet fraction that increases with increasing parton $x$, and thus
predicts a color-singlet fraction that increases with $E_T$ and $\Delta\eta$.
As seen in Fig.~\ref{f:fitnaive},
  the soft-color model (dashed line) 
  gives a reasonable description of the data.

The exchange of
  two gluons in a color-singlet state was originally
  proposed as a simple mechanism to produce rapidity gaps
  between jets~\cite{Dok,BJ} with a predicted color-singlet fraction 
  on the order of $1\%$ ~\cite{BJ}.
Within the na\"{\i}ve color-factor formalism,
  the color factors for this simple two-gluon model
  are $C_{qq}\!=\!1$, $C_{qg}\!=\!9/4$, and $C_{gg}\!=\!(9/4)^2$~\cite{BJ}.
Since the coupling to gluons is stronger for the
two-gluon singlet than for
  single-gluon exchange,  the
  observed color-singlet fraction is 
  expected to decrease with increasing $x$ due to 
  the parton distributions, in contrast with the soft-color model.
As seen in Fig.~\ref{f:fitnaive}, 
  the ``two-gluon'' model (dot-dashed line) does not give an adequate 
  description of the data.

We also fit a ``single-gluon'' model in which the color singlet 
  is indistinguishable from single-gluon exchange, except in
  production rate.
In this model, $C_{qq}\!=\!C_{qg}\!=\!C_{gg}\!=\!1$ and 
  there is no dependence on $E_T$ or $\Delta\eta$ ($f_{S}$ is constant).
The comparison of the single-gluon
  model and data is shown in Fig.~\ref{f:fitnaive}.
The single-gluon
  model (dotted line) does not reproduce
the rising trend of the data with $E_T$ or $\Delta\eta$.

To determine the best fit within the color-factor formalism, we 
  fit a ``free-factor'' model in which 
  the three color factors are allowed to float ($f_{N} \equiv 1$).
We obtain the following values for the color factors:
$C_{qq}=0.025 ^{+0.001}_{-0.008}$, $C_{qg}=0.001^{+0.003}_{-0.001}$,
and $C_{gg}=0^{+0.007}_{-0}$. The best fit to the
data, shown as the solid line in Fig.~\ref{f:fitnaive}, 
thus favors a color-singlet
that couples strongly to quarks with only a few per cent contribution from 
the quark-gluon term ($C_{qq}$\,:\,$C_{qg}$\,:\,$C_{gg} =
1$\,:\,$0.04^{}_{}$\,:\,$0^{}_{}$).

Table~\ref{t:fits} shows the relative likelihood 
for each of the color-singlet models, ordered 
by relative likelihood.
The relative likelihood is the likelihood of each model 
normalized to that of the free-factor model,
which provides an excellent description of the data.
Table~\ref{t:fits} also displays
the average value of $\chi^2$ divided by the number of degrees
of freedom ($\langle\chi^2\rangle/df$), which
is calculated by integrating all possible $\chi^2$ values over 
the prior probability distributions
and dividing by the number of data points minus the number of fit parameters.
If the Monte Carlo
expectation were identical to the observed data in each bin then
by definition $\langle\chi^2\rangle = 0$.

The soft-color model
gives a good descriptions of both data variables, although the
  free-factor model has a likelihood that is several times larger
than the soft-color model. 
The color factors for both of these 
  models indicate that the data are 
  consistent with a color-singlet 
  that couples primarily to quarks.
The single-gluon model does not give as good a fit to the data as the
soft-color model, but a color-singlet that couples like a single gluon
cannot be excluded. None of the other models 
adequately describes the color-singlet fraction for both variables.

\begin{table}[hbt]
\vspace{-.1in}
\begin{center}
\begin{tabular}{|l||l|l||l|l|}
  & \multicolumn{2}{l||}{Fit to $f_S^{\twoD}(E_T)$}
  & \multicolumn{2}{l|}{Fit to $f_S^{\twoD}(\Delta\eta)$} \\ \cline{2-5}
Model    & Likelihood  & $\langle\chi^2\rangle/df$
& Likelihood & $\langle\chi^2\rangle/df$ \\ \hline \hline
``free-factor''   & $\equiv 1.0$ &1/5 & $\equiv 1.0$  & 6/9 \\
``soft-color''    & $1/2.3$  &3/7 & $1/6.0$ & 10/11\\
``single-gluon''  & $1/28$  &8/7 & $1/350$ & 18/11 \\
BFKL              & $1/10^{5}$  &23/7   & $1/66$ & 15/11\\
photon            & $1/660$  &14/7 &   $1/10^{5}$ & 30/11 \\
``two-gluon''     & $1/1100$  &15/7 & $1/10^{5}$ & 30/11 \\
BFKL \ttt         & $1/10^{8}$  &38/7  & $1/10^{9}$  & 51/11 \\
U(1)              & $1/10^{14}$ &64/7 &  $ 1/10^{11}$ & 58/11\\ 
\end{tabular}
\caption{Results of fitting the various color-singlet models to the
color-singlet fraction measurements $f^{\twoD}_S$($E_T$) and
$f^{\twoD}_S$($\Delta\eta$).
For each fit, the relative likelihood 
compared to the ``free-factor'' model and the
$\langle\chi^2\rangle/df$ (average $\chi^2$ per degree of freedom)
are shown.
}
\label{t:fits}
\end{center}
\vspace{-.3in}
\end{table}

Since the free-factor and soft-color models are in good agreement with
the $E_T$ and $\Delta\eta$ dependence at $\sqrt{s} = 1800$~GeV, we use
them to calculate $R_{1800}^{630}$.
This results in a predicted value of $R_{1800}^{630}=1.5\pm 0.1$, where
the uncertainty includes the statistical error on the Monte Carlo samples
as well as the difference between the two models. If we assume
that the models are correct, then the discrepancy between the
predicted value and the measured value of $R_{1800}^{630}=3.4 \pm 1.2$
could be attributed to the  $\sqrt{s}$ dependence of the survival probability.
With this assumption, we obtain a value of $2.2 \pm 0.8$
for the ratio of survival probabilities between the 
two energies, which is consistent with a recent estimate  
for this ratio of $2.2\pm 0.2$~\cite{Gotsman}.

In conclusion,
  we have presented new information on 
  the fraction of dijet events
  produced via color-singlet exchange.
For our largest data sample, $E_{T2}\!>\!30$~GeV
at $\sqrt{s} = 1800$~GeV, we measure a color-singlet
fraction $f_S = [0.94\pm0.04$(stat)$\pm 0.12$(syst)]\%.
The measured ratio of color-singlet fractions 
between $\sqrt{s}= 630$~GeV and 1800~GeV
for jets with
$E_{T2}\!>\!12$~GeV is 
$3.4 \pm 1.2$.
The measured color-singlet fraction 
at $\sqrt{s}= 1800$~GeV
tends to increase with dijet $E_T$ and $\Delta\eta$.
Under the assumption of an $x$-independent survival probability,
the data favor soft-color rearrangement models in which the color-singlet is
preferentially exchanged in processes with initial quarks. 
A model in which the color-singlet coupling is similar to single-gluon
exchange cannot be excluded.
Photon-like, U(1), and current two-gluon models
do not adequately describe the data.



We thank Errol Gotsman, Francis Halzen, Hitoshi Murayama, and 
Dieter Zeppenfeld for their comments and useful discussions.
We thank the staffs at Fermilab and collaborating institutions for their
contributions to this work, and acknowledge support from the
Department of Energy and National Science Foundation (U.S.A.),
Commissariat  \` a L'Energie Atomique (France),
Ministry for Science and Technology and Ministry for Atomic
Energy (Russia),
CAPES and CNPq (Brazil),
Departments of Atomic Energy and Science and Education (India),
Colciencias (Colombia),
CONACyT (Mexico),
Ministry of Education and KOSEF (Korea),
and CONICET and UBACyT (Argentina).
%



\begin{references}
\bibitem{vtx} The vertex algorithm was optimized for primary vertex 
resolution.
\bibitem{pdfsNote} 
 The Monte Carlo results are relatively
  insensitive to the choice of parton distribution functions for the range of 
  $x$ in the data samples ($0.08<x<0.42$).
\bibitem{MuellerNote} We used the parameters specified in Ref.~\cite{Mueller}.
\bibitem{photonNote} Photon exchange processes in 
 {\footnotesize HERWIG}
  are subject to similar higher-order effects to those seen in the BFKL
 color-singlet fraction as a function of $\Delta\eta$.
\bibitem  {HalzenNote} Although 
  this approach is similar to one found in Ref.~\cite{Halzen},
  we developed it independently to complement the existing
  color-singlet models in 
 {\footnotesize HERWIG}.
\end{references}
\end{document}